\documentclass[english,russian,12pt,twoside]{article}
\usepackage{amssymb,amsmath}
\usepackage[russian]{babel}
\usepackage[cp866]{inputenc}
\begin{document}
MSC 35Q60, 83C50\vspace{2mm}

\begin{center} {\bf ПРЕОБРАЗОВАНИЯ ЛОРЕНЦА ДЛЯ ОДНОЙ БИКВАТЕРНИОННОЙ МОДЕЛИ
ЭЛЕКТРО-ГРАВИМАГНИТНОГО ПОЛЯ. ЗАКОНЫ СОХРАНЕНИЯ } \vspace{4mm}

{\bf Л.А.Алексеева }
\end{center}

\vspace{2mm}

\centerline{\textit{Институт математики МОН РК, Алматы,
Казахстан}}
\centerline{alexeeva@math.kz }\vspace{6mm}

Рассматривается одна бикватернионная модель
электро-гравимагнитного (ЭГМ) поля, для  построения которой
использовалась комплексная гамильтонова форма  симметризованных
уравнений Максвелла [1].  В [2] показано, что гамильтонова форма
позволяет легко перейти к бикватернионной записи этих уравнений и
законов сохранения. Следует отметить, что свои уравнения Максвелл
дал в кватернионной форме, а ныне принятая и широко используемая
принадлежит Хевисайду [3]. Кватернионные формы уравнений Максвелла
ранее получали  и другие авторы [3-5]. Они различаются в
зависимости от того,  как вводятся кватернионы напряженности,
зарядов и токов ЭМ-поля, а также операции на их алгебрах. Однако
эти формы использовались в основном лишь для исследования решений
уравнений Максвелла. Подобные формы также  использовал  В.В.
Кассандров для построения своей модели  поля [6].

Здесь используется скалярно-векторная запись бикватернионов,
которая очень наглядна и удивительно приспособлена для записи
физических величин и уравнений. Рассмотрена задача Коши для
комплексных градиентов
 в пространстве бикватернионов
и получены их решения.

 С введением бикватерниона
\textit{силы-мощности} развивается бикватернионный подход для
построения уравнений взаимодействия ЭГМ-полей, порождаемых
различными зарядами и токами, и на их основе аналоги трех законов
Ньютона для свободных и взаимодействующих зарядов-токов, а также
суммарного поля взаимодействий.  Рассмотрена связь полученных
уравнений с известными для механики сплошной среды. Получены
законы преобразования и сохранения энергии при  взаимодействии.

Исследована инвариантность уравнений модели ЭГМ-поля при
преобразованиях Лоренца,
 и, в частности, закона сохранения заряда-тока. Показано, что при взаимодействии полей,
этот закон отличается от общеизвестного. Поэтому, в отличие от
ранее предложенной нами модели ЭГМ-поля в работах [7,8], для
замыкания уравнения трансформации зарядов-токов предложена новая
модификация уравнений
  Максвелла с введением скалярного поля в бикватернион напряжености ЭГМ-поля.
Построены релятивистские формулы преобразования плотностей масс и
зарядов, токов, сил и их мощностей.

\vspace{3mm}

{\bf 1. Гамильтонова форма уравнений Максвелла.} Симметризованные
уравнения Максвелла для ЭМ-поля можно записать в виде одного
векторного и одного скалярного уравнения. В пространстве
Минковского ${\bf M}=R^{1+3} = \left\{ {(\tau ,x) = (ct,x_1 ,x_2
,x_3 )} \right\}$ они имеют следующий вид [1]:
\begin{equation}\label{(1.1)}
\partial _\tau  A + i\,rotA + J = 0,
\end{equation}
\begin{equation}\label{(1.2)}
\rho  = div\,A ,
\end{equation}
где $A$ - комплексный вектор напряженности поля:
\begin{equation}\label{(1.3)}
A = A^E  + i\,A^H  = \sqrt \varepsilon \, E + i\,\sqrt \mu \, H,
\end{equation}
$E,H$    -    напряженности   электрического    и магнитного
полей, $\varepsilon,\;\mu $- константы, характеризующие
электрическую проводимость и магнитную   проницаемость среды,  $c
= 1/\sqrt   {\varepsilon  \mu  }  $- скорость  ЭМ-волн. Плотность
заряда $\rho$ и  $J$   - ток выражаются через электрические и
магнитные заряды и токи формулами:
\begin{equation}\label{(1.4)}
\rho  = \rho ^E/\sqrt{\varepsilon}  - i\,
 \rho ^H /\sqrt{\mu}, \quad J = \sqrt {\mu }\, j^E  - i\sqrt \varepsilon \, j^H ,
\end{equation}
\begin{equation}\label{(1.5)}
\rho ^E  = \varepsilon \,div\,E,\quad \rho ^H  =  - \mu \,div\,H.
\end{equation}
Плотность энергия  A-поля  $W$ и   вектор   Пойнтинга $P$
определяются выражениями:
\begin{equation}\label{(1.6)}
W = 0,5\left( {\varepsilon \left\| E \right\|^2  + \mu \left\| H
\right\|^2 } \right) =  0,5(A,\bar A),\quad
 P = c^{ - 1} E \times H =
0,5i\,[A,\bar A],
\end{equation}
 где  $\bar  A = \sqrt \varepsilon  E - i\sqrt
\mu  H$ - комплексно-сопряженное $A$. Здесь всюду
$(a,b)=\sum_{i=1}^3 a_i b_i$, $[a,b] =a \times b=\sum_{i,j,k=1}^3{
\varepsilon _{ijk}e_i a_j b_k} $ -  скалярное и  векторное
произведения $a$ и $b$ соответственно,
 $\varepsilon _{ijk}$ - псевдотензор Леви-Чивита, $e_i$-орты
 декартовой системы координат. Как видим из (6), плотность энергии -
 это просто половина квадрата модуля комплексного вектора А.

В  уравнениях  Максвелла плотность магнитного заряда $\rho ^H  =
0$, т.к. магнитное поле - вихревое:$div\,H=0.$ Известно, что
гравитационное поле является скалярным, описывается скалярным
гравитационным потенциалом, который зависит от распределения масс.
Здесь предлагаем объединить эти два поля в одно --
\textit{гравимагнитное}, что можно сделать введением
гравитационной плотности в уравнения Максвелла. В частности,
предположим, что \textit{плотность  $\rho  ^H $ эквивалентна
плотности гравитационной  массы.} Далее  покажем,  что эта
гипотеза имеет теоретические подтверждения, приводящие к весьма
правдоподобным следствиям.

Отсюда   следует,  что потенциальная часть вектора  $H$ описывает
гравитационное поле, а вихревая - магнитное, поэтому $H$-поле  -
это гравимагнитное поле. Следовательно,  $A$-поле является
\textit{электро-гравимагнитным}. Поскольку его размерность
определяется плотностью энергии, его можно назвать
\emph{энергетическим.}

Будем называть $j^H $ \textit{гравимагнитным } током. При $\rho ^H
= 0$  это  чисто \textit{магнитные} токи, при потенциальном $H$
токи \textit{массовые}.

Заметим,  что все соотношения для A-поля (а не для $E$ и  $H$) не
содержат  констант среды,  в частности,  скорость электромагнитных
волн, которая во введенной системе координат безразмерна и равна
1.

Приведем здесь также некоторые известные утверждения для А-поля,
 которые являются следствием уравнений Максвелла [1].

Т  е  о р е м а  1.1. \textit{При заданных токах и зарядах
решение} \textrm{(\ref{(1.1)})} \textit{является решением
волнового уравнения}:
 \begin{equation}\label{(1.7)}
    \Box \,A  =  (\partial  _\tau  ^2  - \Delta )A = i\,  rot\,J  -
grad\rho  - \partial _\tau  J,    \end{equation} \textit{и
удовлетворяет законам  сохранения заряда и энергии}:
\begin{equation}\label{(1.9)}
    \partial _\tau  \rho  + div\,J = 0,
\end{equation}
\begin{equation}\label{(1.10)}
    \partial _\tau  W + div\,P =  - {\mathop{\rm Re}\nolimits} (J,\bar
                         A) = c^{-1}(j^H H - j^E E).
\end{equation}

Система уравнений Максвелла незамкнута. Она позволяет по заданным
зарядам  и токам определять поле, и наоборот, при заданном поле
находить порождающие его заряды и токи. Если последние неизвестны,
то для ее замыкания обычно используют уравнения механики сплошных
сред. Однако здесь мы поступим иным образом, используя
бикватернионную запись этих уравнений и законы Ньютона.

Для перехода к бикватернионной записи этих и последующих уравнений
дадим краткое описание функционального пространства бикватернионов
и операций над нем.

\vspace{2mm}

 \textbf{ 2. Бикватернионы на М и их комплексные градиенты}.
Рассмотрим функциональное пространство   бикватернионов --  это
пространство комплексных кватернионов:  $K({\bf R}^{1+3}) = \{
{\bf  F} = f(\tau,x ) + F(\tau,x )\} $,  где  $f$ -
комплекснозначная функция , а $F$ - трехмерная вектор-функция с
комплексными компонентами, $f$ и $F$ - локально интегрируемы  и
дифференцируемы на \textbf{M}. Пространство К - ассоциативная, но
некоммутативная алгебра со сложением: ${\bf  F}  +  {\bf  G}
=  (f  +  g)  + (F  + G)$, и операцией кватернионного умножения (
$ \circ $):
\begin{equation}\label{(2.1)}
{\bf F} \circ {\bf G} = (f + F) \circ (g + G) = (fg - (F,G)) +
                 (fG + gF + [F,G]).
\end{equation}
Бикватернион вида ${\bf \bar F}  =  \bar f +\bar F$ называется
\textit{комплексно-сопряженным}, а ${\bf F}^{*}  =  \bar f - \bar
F$ называется \textit{сопряженным}.   Если   ${\bf   F}^*    =
{\bf F}$, бикватернион называется \textit{самосопряженным}.

Скалярным произведением бикватернионов $ {\bf F}_1 ,{\bf F}_2 $
назовем билинейную операцию \\ $ \left( {{\bf F}_1 ,{\bf F}_2 }
\right) = f_1 f_2  + \left( {F_1 ,F_2 } \right) $. Норма
бикватерниона  $ \left\| {\bf F} \right\| = \sqrt {\left( {{\bf
F},{\bf \bar F}} \right)}  = \sqrt {f \cdot \bar f + \left(
{F,\bar F} \right)}  = \sqrt {\left| f \right|^2  + \left\| F
\right\|^2 } $, а псевдонормой бикватерниона   назовем  величину $
\left\langle {\bf F} \right\rangle  = \sqrt {f \cdot \bar f -
\left( {F,\bar F} \right)}  = \sqrt {\left| f \right|^2  - \left\|
F \right\|^2 } $.

Далее   используются   дифференциальные  операторы   --   взаимные
комплексные градиенты: ${\bf D}^ +   = \partial _\tau   + i\nabla
,\quad {\bf D}^ -
                   = \partial _\tau   - i\nabla$ ,
где $\nabla  = grad = (\partial _1 ,\partial _2 ,\partial  _3  )$.
Их действие на К определено как в алгебре кватернионов:
(соответственно знакам)
 $$
   {\bf D}^ \pm  {\bf F} = (\partial _\tau   \pm i\nabla ) \circ (f +
   F) = (\partial _\tau  f \mp i\,(\nabla ,F)) \pm\partial _\tau  F \pm
                      i\nabla f \pm i[\nabla ,F]=
  $$
  $$ = (\partial _\tau  f \mp i\,div\,F) \pm \partial _\tau  F \pm igrad\,f
                            \pm i\,rot\,F.$$
Заметим, что в смысле  выше  данных определений: $({\bf D}^ - )^*
= {\bf D}^ -  ,\;({\bf D}^ +  )^*  = {\bf D}^ +  $. Легко
проверить, что волновой оператор представим в виде:
$$ \Box=\frac{\partial^2}{\partial\tau^2}-\triangle= {\bf D}^ -   \circ {\bf D}^ +
= {\bf D}^ +   \circ {\bf D}^ -  .$$
 Используя это свойство, можно
строить частные решения дифференциальных уравнений на
\emph{К}(\textbf{М}) вида:
\begin{equation}\label{(2.2)}
 {\bf D}^ \pm  {\bf K} = {\bf G}.
\end{equation}
Отсюда следует, что $\Box {\bf K} = {\bf D}^ \mp {\bf G}$, его
решением является следующая свертка (*)
\begin{equation}\label{(2.3)}
{\bf K} = {\bf D}^ \mp  {\bf G} * \psi,
\end{equation}
где   $\psi (\tau ,x)$ - фундаментальное решение волнового
уравнения: $\Box\psi  = \delta (\tau) \delta(x).$ Это решение
является также и решением (\ref{(2.2)}). Действительно, используя
свойство дифференцирования свертки,  получим
\[
{\bf D}^ \pm  {\bf K} = {\bf D}^ \pm  {\bf D}^ \mp  \left( {{\bf
G} * } \right.\left. \psi  \right) = \Box\left( {{\bf G} * }
\right.\left. \psi  \right) = \left( {{\bf G} * } \right.\left.
\Box{\psi } \right) = {\bf G*}\delta (\tau )\delta (x) = {\bf G}.
\]

Фундаментальные решения определяются с точностью до решений
однородного волнового уравнения. Для задач с начальными по времени
условиями в качестве фундаментального решения удобно использовать
простой слой на световом конусе $\tau  = \left\| x \right\|$:
$$\psi  = (4\pi \left\| x \right\|)^{ - 1}
\delta (\tau  - \left\| x \right\|),$$ которую назовем
\emph{волновой функцией}.

В этом случае, как легко показать, записав свертку в интегральном
виде, решение (\ref{(2.3)}) будет равно нулю при $\tau=0$.
Воспользуемся им для построения решений уравнения (13) с  данными
Коши.

\emph{Задача Коши.} Пусть известны начальные условия: ${\bf
K}(0,x)={\bf K_0}(x).$ Требуется построить решение
уравнения(\ref{(2.2)}), удовлетворяющее этим данным.

Используем для этого аппарат теории обобщенных функций [9 ].
Рассмотрим регулярные обобщенные функции вида
 ${\bf \widehat{G}}=H(\tau){\bf G}(\tau,x)$, где $H(\tau)$- функция Хевисайда.
 Используя  дифференцирование обобщенных функций (${\bf \widehat D}^ \pm $), получим
  ${\bf \widehat D}^ \pm  {\bf \widehat{K}} = {\bf \widehat{G}}+\delta(\tau) {\bf {K_0}}(x)$.
Следовательно,
\begin{equation}\label{(2.4)}
{\bf H(\tau)K}(\tau,x) = {\bf D}^ \mp  \{H(\tau){\bf {G}} *
\psi\}+{\bf G}(0,x)
 \mathop *\limits_x \psi+{\bf D}^ \mp\{{\bf K_0}(x) \mathop *\limits_x \psi\}
 \end{equation}
(здесь знак "$\mathop *\limits_x$ " означает, что свертка берется
только по $x$). Эта формула является обобщением формулы Кирхгофа
для решения задачи Коши для волнового уравнения [ 9]. Ее
интегральная запись легко выписывается, с учетом вида полной и
неполной свертки с $\psi$. А именно,
\begin{equation}\label{(2.5)}
{4\pi }{\textbf{K}}(\tau,x) =- {\bf D}^ \mp  \left\{\int\limits_{r
\le \tau } {\frac{{{\bf G}(\tau  - r,y)}}{r}} dV(y)
+\tau^{-1}\int\limits_{r =\tau } {\bf K_0}(y)
dS(y)\right\}-\tau^{-1} \int\limits_{r =\tau } {\bf G}(0,y) dS(y),
 \end{equation}
где $ r=\|y-x\|,\,dV(y)=dy_1dy_2dy_3$,  $dS(y)$ - дифференциал
площади сферы.

Перейдем к бикватернионному
 представлению уравнений А-поля [2].
 \vspace{2mm}

 \textbf{ 3. Бикватернионы А-поля}.
Вводятся бикватернионы : \textit{потенциал} ${\bf \Phi } = i\phi -
\Psi $, \textit{напряженность}  $\textbf{A}=0+A$,
\textit{плотность заряда-тока}${\bf \Theta } = -i\rho - J$,
\textit{плотность энергии-импульса} ${\bf \Xi }= 0,5\,{\bf A}^*
\circ {\bf A}  = W + iP$ .

Уравнения Максвелла (1)-(2) в пространстве бикватернионов имеют
простой вид:
\begin{equation}\label{(3.1)}
{\bf D}^+  {\bf A} = {\bf \Theta } .
\end{equation}
Если  потенциал  удовлетворяет лоренцевой калибровке:
$\partial_\tau  \phi  - div\,\Psi  = 0$, то
$${\bf A} = {\bf D}^ -  {\bf \Phi } .$$
Откуда, взяв соответствующий комплексный градиент, получаем
волновые уравнения:
\begin{equation}\label{(3.2)}
\Box{\bf \Phi } =   {\bf\Theta },
\end{equation}
\begin{equation}\label{(3.3)}
 \Box {\bf A} =  {\bf D}^- {\bf \Theta }.
\end{equation}

Отсюда следует, что просто последовательное (тройное) взятие
комплексных градиентов от потенциала А-поля, определяет
бикватернионы, соответствующие напряженности поля, зарядам и
токам. Скалярная часть комплексного градиента кватерниона
энергии-импульса А-поля дает закон сохранения энергии [2].

Итак,  заряды и токи - это просто \textit{физическое проявление
комплексного градиента напряженности ЭГМ-поля}. \vspace{2mm}

\emph{Задача Коши для уравнения Максвелла }. Как
следует из уравнения (\ref{(2.4)}), при известных зарядах-токах и
начальных данных ${\bf A}(0,x)={\bf A_0}(x)$,
  решение (\ref{(3.1)}) дается формулой:
\begin{equation}\label{(3.4)}
{4\pi }{\textbf{A}} =- {\bf D}^ -  \left\{\int\limits_{r \le \tau
} {\frac{{{\bf \Theta}(\tau  - r,y)}}{r}} dV(y) +
\tau^{-1}\int\limits_{r =\tau } {\bf A_0}(y) dS(y)\right\} -
\tau^{-1} \int\limits_{r =\tau } {\bf \Theta}(0,y) dS(y).
 \end{equation}
Отсюда легко записать интегральные представления для векторов напряженности ЭГМ-поля \emph{E,H}.
\vspace{3mm}

\textbf{4.  Преобразование Лоренца К на М.} Преобразования Лоренца
бикватернионов на пространстве Минковского
 удобно строить, используя алгебру кватернионов. Для этой цели
 кватернизируем \textbf{М}, вводя комплексно-сопряженные бикватернионы: $
{\bf Z} = \tau  + ix$, ${\bf \bar Z} = \tau  - ix $.  Легко
видеть, что
\[
{\rm   }{\bf Z} = {\bf Z}^* ,\,\,{\bf \bar Z} = {\bf \bar Z}^*
,\,\,\left\| {\bf Z} \right\|^2  = \left\| {{\bf \bar Z}}
\right\|^2 = ({\bf Z},{\bf \bar Z}),\,\,\left\langle {\bf Z}
\right\rangle ^2  = \left\langle {{\bf \bar Z}} \right\rangle ^2 =
{\bf Z} \circ {\bf \bar Z}.
\]
Введем самосопряженные бикватернионы $ {\bf U} = ch\theta  +
iesh\theta ,\,\,{\bf \bar U}  = ch\theta  - iesh\theta ,\quad
\left\| e \right\| = 1$,  $\theta$ -действительное число, ${\bf
U}\circ{\bf \bar U}=1 $.
 Прямым вычислением доказываются следующие леммы.

   \textbf{ Л е м м а  4.1. } \emph{Классическое преобразование Лоренца} $
L:{\bf {\rm Z}} \to {\bf {\rm Z}}^{\bf '} $ \emph{имеет вид}:   $$
{\bf Z}' = {\bf U} \circ {\bf Z} \circ {\bf U},\quad {\bf Z} =
{\bf \bar U}
  \circ {\bf Z}' \circ {\bf \bar U},$$

Если вести  обозначения: $ ch2\theta  = \frac{1}{{\sqrt {1 - v^2 }
}},\quad sh2\theta  = \frac{v}{{\sqrt {1 - v^2 } }},\quad\left| v
\right| < 1 ,$
 то скалярная
и векторная часть бикватернионов запишется в  виде известных
релятивистских формул:
\[
\tau ' = \frac{{\tau  + v(e,x)}}{{\sqrt {1 - v^2 } }},\quad x' =
(x - e(e,x)) + e\frac{{(e,x) + v\tau }}{{\sqrt {1 - v^2 } }},
\]
\[
\tau  = \frac{{\tau ' - v(e,x)}}{{\sqrt {1 - v^2 } }},\quad x =
(x' - e(e,x')) + e\frac{{(e,x') - v\tau '}}{{\sqrt {1 - v^2 } }},
\]
что соответствует движению системы координат X в направлении
вектора \emph{е }с безразмерной скоростью \emph{v}. Легко видеть,
что сохраняется псевдонорма:  \[ \left\langle {{\bf Z'}}
\right\rangle ^{\bf 2}  = {\bf U} \circ {\bf Z} \circ {\bf U}
\circ {\bf \bar U} \circ {\bf \bar Z} \circ {\bf \bar U} =
\left\langle {\bf Z} \right\rangle ^{\bf 2}.
\]

\textbf{ Лемма 4.2.} \emph{Сопряженные  кватернионы} $ {\bf W} =
\cos \varphi  + e\sin \varphi ,\,\, {\bf W}^{\bf *}  = \cos
\varphi  - e\sin \varphi ,$\\$ \left\| e \right\| = 1 $,
\emph{определяют группу преобразований на \textbf{\textrm{М}},
ортогональных на векторной части Z} : $ {\bf Z}^{\bf '}  = {\bf W}
\circ {\bf Z} \circ {\bf W}^{\bf *} ,\,\,\, {\bf Z} = {\bf W}^{\bf
*}  \circ {\bf Z}^{\bf '}  \circ {\bf W} $.

  Это преобразование есть вращение вокруг  вектора  \emph{e}  на угол $2\varphi$ .
    Следствием этих двух лемм является

\textbf{Лемма 4.3.} \emph{Преобразование Лоренца на
\textbf{\textrm{М}} можно определить как преобразование вида}:
\begin{equation}\label{(4.1)}
{\bf Z}^{\bf '}  = {\bf L} \circ {\bf Z} \circ {\bf L}^* ,\,\,
{\bf Z} = {\bf L}^*  \circ {\bf Z}^{\bf '}  \circ {\bf L},
\end{equation}
\emph{где} ${\bf L} = {\bf W} \circ {\bf U} = ch(\theta  +i\varphi
) + iesh(\theta  +i\varphi ) ,\,\,\, {\bf L}^*  = {\bf U}^*  \circ
{\bf W}^*  = ch(\theta  - i\varphi ) + iesh(\theta  -i\varphi ).$
\emph{При этом
}$\langle\textbf{Z}\rangle=\langle\textbf{Z}'\rangle$.

Легко видеть, что ${\bf \bar L} \circ {\bf L}^*  = {\bf L}^* \circ
{\bf \bar L} = 1 $, поэтому псевдонорма  $\textbf{Z}$сохраняется.

Взаимные комплексные градиенты при преобразованиях Лоренца L
 преобразуются  в соответствии со следующей  леммой.

\textbf{
    Л е м м а  4.4.} \emph{Если} $
{\bf Z}^{\bf '}  = {\bf L} \circ {\bf Z} \circ {\bf L}^* $,
\emph{то} $ {\bf D}' = {\bf \bar L}^*  \circ {\bf D} \circ {\bf
L},\quad {\bf D} = {\bf L} \circ {\bf D}' \circ {\bf\bar L}^* ,$
\emph{где} $ {\bf D} = {\bf D}^ + $ \emph{или}  $ {\bf D} = {\bf
D}^ - .$

На основе этой леммы рассмотрим, как меняется уравнение типа
(\ref{(2.2)}) при преобразовании Лоренца.

  \textbf{  Т е о р е м а  4.1.} \emph{При действии преобразования Лоренца на}
   \textbf{M } \emph{сохраняется вид уравнения}:
\[
\left( {\frac{\partial }{{\partial \tau '}} \pm i\nabla '}
\right){\bf K'} = {\bf G'},
\]
\emph{где}      $ {\bf K}^{\bf '}  = {\bf \bar L}^*  \circ {\bf K}
\circ {\bf L} ,\,\,\, {\bf G}^{\bf '}  = {\bf \bar L}^*  \circ
{\bf G} \circ {\bf L}. $

Действительно, используя ассоциативность произведения и свойства
\textbf{L},  получим
\[
{\bf D}'{\bf K}' = \left( {{\bf \bar L}^*  \circ {\bf D} \circ
{\bf L}} \right)\left( {{\bf \bar L}^*  \circ {\bf K} \circ {\bf
L}} \right) = {\bf \bar L}^*  \circ {\bf D} \circ {\bf K} \circ
{\bf L} = {\bf \bar L}^* \circ {\bf G} \circ {\bf L} = {\bf G'}.
\]
Ч.т.д.

Следовательно преобразования Лоренца для уравнения  Максвелла
имеют вид:
$$\textbf{D}^+{\bf A'} = {\bf \Theta'},\,\,
\emph{где}      \,\, {\bf A}^{\bf '}  = {\bf \bar L}^*  \circ {\bf
A} \circ {\bf L} ,\,\,\, {\bf \Theta}^{\bf '}  = {\bf \bar L}^*
\circ {\bf \Theta} \circ {\bf L}. $$ Расписывая эти формулы  в
подвижной системе координат ($\varphi=0$), получим

\emph{Релятивистские формулы для напряженности, зарядов и токов}:
\begin{equation}\label{(4.2)}
A' =(A-e(e,A)) + e{\frac{{(e,A)  }}{{\sqrt {1 - v^2 } }}}
\end{equation}

\begin{equation}\label{(4.3)}
\rho' = \frac{{\rho - v(e,J)}}{{\sqrt {1 - v^2 } }},\quad  J' =
(J-e(e,J)) + e{\frac{{(e,J) - v\rho }}{{\sqrt {1 - v^2 } }}}
\end{equation}
Как видим, напряженность А-поля здесь всегда увеличивается в
направлении вектора \emph{е}. В отсутствие токов, происходит
увеличение заряда-массы. При  наличии токов, в зависимости от
направления их движения, заряд-масса может как увеличиваться, так
и уменьшаться.

\vspace{2mm}

 \textbf{ 5.  Третий закон Ньютона. Мощность и плотность объемных сил.}
   Рассмотрим  два ЭГМ-поля ${\bf A}$ и ${\bf
A}'$, ${\bf \Theta  }$ и ${\bf \Theta'}$  -- соответствующие им
(или порождающие их) заряды-токи.  Назовем бикватернион
\begin{equation}\label{(5.1)}
{\bf F} = M - iF =   \;{\bf \Theta } \circ {\bf A}' =  -
                (i\rho  + J) \circ A' = (A',J) - i\rho A' + [A',J]
\end{equation}
   \textit{плотностью мощности-силы}, действующей  со  стороны
поля $A'$ на заряды  и  токи  поля $A$ .   Действительно,  с
учетом (\ref{(1.3)}),(\ref{(1.4)}), скалярная часть имеет вид
плотности мощности действующих сил:
\begin{equation}\label{(5.2)}
 M = (A',J) = c^{ - 1} ((E',j^E ) + (H',j^H )) + i((B',j^E ) -
 (D',j^H ))
\end{equation}
Выделяя  действительную  и мнимую части  векторной  составляющей
бикватерниона, получим выражения для плотности объемных сил
$\left( {F = F^H  + i\,F^E } \right)$:
\begin{equation}\label{(5.3)}
 F^H= \rho ^E E' + \rho ^H H' + j^E\times B' - j^H  \times D'
\end{equation}
\begin{equation}\label{(5.4)}
 F^E = c\left( {\rho ^E B' - \rho ^H D'} \right) + c^{ - 1}
\left( {E' \times j^E  + H' \times j^H } \right)
\end{equation}
Здесь  $B  =  \mu H$- аналог вектора магнитной индукции (в
вихревой части совпадает с ним), $ D = \varepsilon E$ -вектор
электрического смещения.

Напряженность  гравитационного  поля  описывается  потенциальной
частью вектора $H$, а роторная часть этого вектора описывает
магнитное поле. Тогда скалярная часть ${\bf \Theta },\,{\bf \Theta
'}$ содержит  плотности  электрического заряда и массы,  а
векторная  - плотности  электрического  тока и тока  массы
(количество  движения массы).

Исходя  из этих предположений, в формуле (\ref{(5.3)}) стоят
известные массовые  силы,  последовательно: кулоновская  сила
$\rho ^E E'$, гравитационная сила $\rho ^H H'$ (точнее совпадает с
ней в потенциальной части \textit{Н'}),
  сила  Лоренца  $j^E\times B'$ (точнее совпадает с ней в вихревой части \textit{В'})) и  новая сила $-D' \times j^H
$, которую  назовем \textit{электромассовой}.  В  действительной
части мощности   (\ref{(5.2)})   стоит мощность  кулоновских,
гравитационных  и магнитных  сил.  Мощность электромассовой силы в
действительную часть (\ref{(5.2)})  не  входит,  т.к. она не
работает на перемещениях массы, поскольку перпендикулярна  ее
скорости. Интересно,  что мощность силы Лоренца  в  действительную
часть (\ref{(5.2)}) также не входит, что свидетельствует в пользу
того, что эта сила перпендикулярна скорости массы, хотя
непосредственно из уравнений Максвелла это не следует.

Естественно,  по  аналогии,  предположить, что  уравнения
(\ref{(5.4)}) описывают    силы,   вызывающие   изменение
электрических токов (электрические  силы), а в мнимой части
\textit{M} стоят соответствующие им мощности.

В    силу    третьего   закона   Ньютона   о    действующих    и
противодействующих силах, предположим, что  должно выполняться:
${\bf F'} =  - {\bf F}.$ Отсюда получим

\textit{  Закон о действии и противодействии  полей }
\begin{equation}\label{(5.5)}
{\bf \Theta } \circ {\bf A'} =  - {\bf \Theta '} \circ {\bf A}.
\end{equation}
Интересно, что в скалярной части он требует равенства плотностей
мощностей соответствующих сил, действующих на заряды и токи
другого поля,  т.е. подобен известному в механике сплошных средств
тождеству взаимности Бетти, которое обычно записывается для работы
сил. \vspace{3mm}

 \textbf{  6. Второй закон Ньютона. Уравнение
трансформации}.    Поле  зарядов  и   токов меняется под
воздействием  поля  других зарядов  и  токов. Как известно,
направление наиболее интенсивного изменения  скалярного поля
описывает его градиент. По аналогии предположим, что изменение
поля зарядов-токов происходит наиболее интенсивно, условно говоря,
в направлении его комплексного градиента. Естественно
предположить,  что  это изменение   должно происходить в
направле\-нии  мощ\-нос\-ти-силы, действующей  со стороны второго
поля на первое.  Поэтому закон изменения заряда-тока поля под
действием другого, подобный второму закону Ньютона, предложен в
виде следующих  уравнений [7,8].

\textit{Уравнения взаимодействия зарядов-токов (полей):}
\begin{equation}\label{(6.1)}
\kappa {\bf D}^ -  {\bf \Theta } = {\bf F} \equiv   {\bf \Theta
 } \circ {\bf A}',\,\,\,\,
\kappa {\bf D}^ -  {\bf \Theta }' =  {\bf \Theta }' \circ {\bf A},
\end{equation}
\begin{equation}\label{(6.2)}
 {\bf \Theta } \circ {\bf A}' =  - {\bf \Theta }' \circ {\bf A},
\end{equation}
\begin{equation}\label{(6.3)}
{\bf D}^ +  {\bf A} ={\bf \Theta } ,
                              \,\,\,\,
{\bf D}^ +  {\bf A}' = {\bf \Theta }' .
\end{equation}
Здесь   уравнения   (\ref{(6.1)}) соответствуют  второму закону
Ньютона, записанному для зарядов-токов каждого из
взаимодействующих полей, а уравнение (\ref{(6.2)})  -  третьему.
Вместе с уравнениями  Максвелла для  этих полей (\ref{(6.3)})  они
дают замкнутую  систему  нелинейных дифференциальных уравнений
для определения ${\bf A, A',\Theta,\Theta'}$ .     Введение
константы взаимодействия $\kappa $ связано с размерностью.
Раскрывая скалярную и векторную часть (\ref{(6.1)}), запишем

\textit{Уравнения трансформации зарядов-токов А-поля }:
\begin{equation}\label{(6.4)}
 i\,\kappa \left( {\partial _\tau  \rho  +div\,J} \right) = M,
\end{equation}
\begin{equation}\label{(6.5)}
i\,\kappa \left( {\partial _\tau  J -i\,rot\,J + \nabla \rho }
  \right) = F.
\end{equation}
Рассмотрим вначале второе уравнение.    С учетом (\ref{(1.2)}),
(\ref{(1.3)}), (\ref{(1.4)}), получим

\textit{Аналог второго закона Ньютона для зарядов-токов:}
\begin{equation}\label{(6.6)}
\kappa  \left(\sqrt \varepsilon\,\partial _\tau  j^H  +\sqrt \mu
 \, rot\,  j^E+\mu  ^{  - 0,5} grad\,\rho  ^H
\right)=\rho ^E E' + \rho ^H H' + j^E  \times B' - j^H \times D',
\end{equation}
\begin{equation}\label{(6.7)}
 \kappa  \left(   \sqrt
\mu\,\partial _\tau  j^E  -\sqrt \varepsilon  \, rot\,j^H
+\varepsilon ^{ - 0,5}  grad\,\rho  ^E \right)= c\left( {\rho ^E
B' - \rho ^H D'} \right) + c^{ - 1} \left( {E' \times j^E  + H'
\times j^H } \right).
\end{equation}
Аналогом  количества движения массы здесь в (\ref{(6.6)}) является
$\kappa \sqrt  \varepsilon   j^H  $. Уравнение (\ref{(6.7)})
описывает воздействие внешнего поля на электрические токи, его
аналог автору неизвестен.

Если одно поле намного сильнее второго, например, если   $W'  >
>  W$,  то  можно
изменением  второго поля под воздействием зарядов  и  токов
первого пренебречь.  В этом случае получаем замкнутую систему
уравнений  для определения  движения зарядов и токов первого поля
под  воздействием зарядов и токов второго: $\kappa {\bf D}^ -
{\bf \Theta }  - {\bf \Theta } \circ {\bf A}'=0$, где   ${\bf
A}'$  известно. Соответствующее им А-поле определяется уравнениями
Максвелла .

Рассмотрим первое уравнение (\ref{(6.4)}). Очевидно, это закон сохранения для
зарядов-токов, который в правой части содержит мощность внешних
действующих сил \emph{М}. Только при \emph{М=}0, например, в
отсутствии внешних полей и соответствующих им зарядов и токов,
справа будет стоять 0. Тогда этот закон принимает хорошо известный
вид:$\partial _\tau  \rho  + div\,J = 0,$ который следовал из
уравнений Максвелла (см. теорема 1.1.).

Значит,
 при взаимодействии ЭГМ-полей уравнения Максвелла несколько меняют вид, а именно появляется скалярная часть
 у бикватерниона напряженности поля:$${\bf A}=ia(\tau,x)+A(\tau,x).$$
Из системы уравнений (\ref{(6.1)}) -(\ref{(6.3)}) следует, что
\begin{equation}\label{(6.8)}
 \Box{\bf A} ={\bf D}^ -{\bf \Theta } =\kappa^{-1}{\bf F}.
\end{equation}
Откуда имеем
\begin{equation}\label{(6.9)}
 -i\,\kappa  \Box \,a =M .
\end{equation}

Заметим, что в системе уравнений Максвелла (1)-(2), первое
уравнение определяет токи,  второе уравнение
 является определением заряда, а закон сохранения заряда
является следствием этих двух уравнений. Его получаем, взяв
дивиргенцию в (1) с учетом (2). Однако последовательный
бикватернионный подход, как здесь показано, приводит к модификации
системы уравнений Максвелла, которая, как следует из
(\ref{(6.3)}), имеет следующий вид:

\emph{Модифицированные уравнения Максвелла}
\begin{equation}\label{(6.10)}
 J=-\partial _\tau  A - i\,rotA +grad \,a  ,
\end{equation}
\begin{equation}\label{(6.11)}
\rho  = div\,A -\partial _\tau a,
\end{equation}
Если заряды и токи известны, эта система уравнений
 для определения $a$ и $A$ замкнута.
Только в замкнутых системах (при отсутствии внешних полей) $a=0$ и
она приобретает вид (\ref{(1.1)})-(\ref{(1.2)}).

\emph{Задача Коши для уравнения трансформации. }Используя формулу
(\ref{(2.3)}), получим
\begin{equation}\label{(6.12)}
\kappa{\bf \Theta}(\tau,x) = {\bf D}^ {+} \{ H(\tau){\bf
{F}}(\tau,x) * \psi\}+{\bf {F }}(0,x) \mathop *\limits_x
\psi+\kappa{\bf D}^ {+}\{{\bf {\Theta}}(0,x) \mathop *\limits_x
\psi\}
 \end{equation}
Уравнения дают систему интегральных уравнений для определения
${\bf \Theta}$, поскольку правая часть содержит  ${\bf \Theta}$ в
\textbf{F}. \vspace{3mm}

\emph{{ Преобразования Лоренца уравнения трансформации. }} (Здесь
штрих означает координаты в подвижной системе координат.) Согласно
теореме 4.1, преобразования
 Лоренца для $\textbf{A}, \,{\bf \Theta },\, \textbf{F}
$  имеют следующий вид:
\begin{equation}\label{(6.13)}
 {\bf A}^{\bf '}  = {\bf \bar L}^*  \circ {\bf A} \circ {\bf
L} ,\,\,\, {{\bf \Theta }}^{\bf '}  = {\bf \bar L}^*  \circ \bf
\Theta  \circ {\bf L},\,\,\, {{\bf {F}}}^{\bf '}  = {\bf \bar L}^*
\circ {{\bf \textsc{F}}} \circ {\bf L},
\end{equation}
Заметим, что преобразование Лоренца для  для мощности-силы
взаимодействия двух полей вида
 (\ref{(5.1)}) имеет тот же вид:
$$
{\bf F'} = {\bf \Theta'_1 } \circ {\bf A'_2} ={\bf \bar L}^* \circ
{\bf \Theta_1}  \circ {\bf L}\circ {\bf \bar L}^*  \circ {\bf
A_2}\circ {\bf L}={\bf \bar L}^*  \circ {\bf \Theta_1}\circ
   {\bf A_2}\circ {\bf L}={\bf \bar L}^*  \circ {\bf F}\circ {\bf L}.
$$
Для $\varphi=0$ соотношения (\ref{(6.13)}) эквивалентны равенствам
(\ref{(4.2)})-(\ref{(4.3)}) и
$${\bf F'} =  (M ch2 \theta-(e,F)sh2 \theta )+i\{F+2e(e,F)sh^2 \theta-M e sh2 \theta \}\Rightarrow$$

\emph{Релятивистские формулы для мощности и силы}:
\begin{equation}\label{(6.14)}
M' = \frac{{M  +v(e,F)}}{{\sqrt {1 - v^2 } }},\quad F' = (F-
e(e,F)) + e{\frac{{(e,F)-vM }}{\sqrt {1 - v^2 } }}.
\end{equation}
Итак мощность также зависит от скорости системы координат. И если
в исходной системе она равна нулю, то в другой будет равна нулю
только в отсутствии внешних сил $F=0$. Поэтому постулировать закон
сохранения заряда в традиционном виде (\ref{(1.9)}) для открытых
систем (систем, подверженных внешним воздействиям), нельзя \footnote [1]{В [7,8]
для сохранения этого закона было введено предположение:M=0.
 Как здесь показано, это было неверное предположение.}.
\vspace{3mm}

\textbf{ 8. Первый закон Ньютона. Свободное поле. } Рассмотрим
А-поле, порождаемое $\bf{\Theta}$, \,в отсутствии других
зарядов-токов.  Назовем  такое поле \textit{свободным}.

В этом случае ${\bf F}=0$, поэтому аналогом  первого закона
Ньютона об инерции массы  в  отсутствии действующих  на  нее  сил
здесь, как следует из  (\ref{(6.1)}) , естественно принять

\textit{    Закон инерции для зарядов-токов А-поля:}
\begin{equation}\label{(8.1)}
  {\bf D}^ -  {\bf \Theta } = {\bf 0},
\end{equation}
что эквивалентно равенствам: $${\partial _\tau  \rho  +div\,J}  =
0,\,\,\,\,
      \partial _\tau  J -i\,\, rot\,J+\nabla \rho   = 0,
$$ или для исходных величин:
\begin{equation}\label{(8.2)}
  \partial _t \rho ^E  + div\,j^E  = 0,\quad \partial _\tau  j^E
          = \sqrt {\varepsilon /\mu }\, rot\,j^H  - c\,grad\,\rho ^E,
\end{equation}
\begin{equation}\label{(8.3)}
\partial _t \rho ^H  + div\,j^H  = 0,\quad \partial _\tau  j^H
        =   -\sqrt {\mu /\varepsilon }\, rot\,j^E  -c\,grad\,\rho ^H.
\end{equation}
Следовательно,  закон сохранения заряда в виде (\ref{(1.9)})
выполняется в отсутствии внешних полей. В этом случае решение
задачи Коши имеет вид:
\begin{equation}\label{(8.4)}
\kappa{\bf \Theta}(\tau,x) = \kappa{\bf D}^ {-}\{{\bf
{\Theta}}_0(x) \mathop *\limits_x \psi\}= - \frac{\kappa
H(\tau)}{4\pi } {\bf D}^ {-}\left\{ \tau^{-1}\int\limits_{r =\tau
} {\bf \Theta}_0(y) dS(y)\right\},
 \end{equation}
а напряженность А-поля определяется соотношениями (\ref{(3.4)}).

\vspace{3mm}

\textbf{10. Первое  начало  термодинамики.} Аналогично плотности
энергии-импульса  А-поля введем плотность энергии-импульса  поля
зарядов-токов:
\begin{equation}\label{(10.1)}
 0,5{\bf \Theta } \circ {\bf \Theta }^*  = \left( {
{\frac{\left\|{\rho _E }\right\|^2}{\varepsilon }}   +
{\frac{\left\|{\rho _H  }\right\|^2 }{\mu }}  + Q} \right) +
i\left( {P_J  - \sqrt {\frac{\mu }{\varepsilon }} \rho ^E j^E  -
\sqrt {\frac{\varepsilon}{\mu }} \rho ^H j^H } \right),
\end{equation}
которая содержит плотность энергии токов:
                                 \[
     Q  = 0,5\left\| J \right\|^2  = 0,5\left( {\mu \left\| {j^E }
   \right\|^2  + \varepsilon \left\| {j^H } \right\|^2 } \right),
                                 \]
где  первое  слагаемое   включает джоулево  тепло  $\left\|  {j^E
} \right\|^2  $,  а  второе - плотность кинетической энергии
массовых токов  $\left\| {j^H } \right\|^2 $, но не только, т.к. в
них входит и энергия вихревой части токов (магнитных токов). Здесь
также введен вектор $P_J  $, подобный вектору Пойнтинга,  но  для
токов:
                                 \[
      P_J  = 0,5i\,J \times \bar J = c^{ - 1} \left[ {j^H ,j^E } \right]
                                 \]
Если гравимагнитный  и электрический токи параллельны, либо один
из них  отсутствует (нулевой), то $P_J  = 0$.  В общем случае $P_J
\ne 0$.

Умножим  скалярно  уравнение (\ref{(6.5)})  на  $  -  i\bar J  $,
сложим  с соответствующим комплексно-сопряженным и поделим на 2. В
результате получим

\textit{Закон сохранения энергии зарядов-токов $\bf{\Theta}$-поля:}
\begin{equation}\label{(10.2)}
 \kappa \left( {\partial _\tau  Q  -\,div\,P_J  +{\mathop{\rm
     Re}\nolimits} \left( {\nabla \rho ,\bar J  } \right)} \right) =
  {\mathop{\rm Im}\nolimits} \left( {F,\bar J  } \right) =c^{-1} \left( ({F^H
             ,j^H } ) + ( {F^E ,j^E }) \right) ,
 \end{equation}
аналогичный  закону  сохранения энергии для А-поля  (теорема
1.1.). Однако  в левой части появилось третье слагаемое.  Нетрудно
видеть, что  этот  закон подобен первому началу термодинамики.
Здесь  второй и третий член в левой части обозначим $-U$. Функция
                                 \[
      U=\,div\,P_J  - \sqrt {\mu /\varepsilon } \left(
  {\nabla \rho ^E ,j^E } \right) -\sqrt {\varepsilon/\mu  } \left(
               {\nabla \rho ^H ,j^H } \right) \]
характеризует собственную скорость изменения плотности энергии
токов $\bf{\Theta}$-поля. Правая часть (\ref{(10.2)}), зависящая
от мощности действующих внешних  сил, может увеличивать или
уменьшать эту скорость.

Для свободного поля первое начало термодинамики  имеет вид: $$
\partial _{\tau} Q  =U.$$

Интегрируя (\ref{(10.2)}) по пространственно-временному цилиндру $\{(D^-+D)\times(0,t)\}$ и используя формулу
 Остро\-град\-ско\-го-Гаусса, получим  интегральное представление первого начала термодинамики:

$$\int\limits_{D^ -  } {\left( {Q (x,t) - Q (x,0)} \right)}
    dV(x) =\int\limits_0^t {dt} \int\limits_D {(P_J ,n)} \,dD(x)- $$
$$- \int\limits_0^t {dt} \int\limits_{D^ -  } {\{\varepsilon^{-1}
\left({\nabla \rho ^E ,j^E } \right)+ \mu^{-1}(\nabla \rho ^H ,j^H
)\, \} dV(x)}+ $$
$$+ c^{-1}\int\limits_0^t {dt} \int\limits_{D^ -  } \left\{(F^H,j^H ) +
 (F^E ,j^E) \right\} dV(x).$$
Здесь $n(x)$ - вектор единичной нормали к границе $D$ открытой области $D^-$ в
$R^3$.

\vspace{3mm}

\textbf{11. Уравнения суммарного поля и энергия взаимодействий}.
Если есть несколько (N) взаимодействующих полей, порождаемых
различными зарядами и токами, то уравнения (\ref{(6.1)}) примут
вид:
\begin{equation}\label{(11.1)}
\kappa {\bf D}^ + {\bf \Theta }^k  + {\bf \Theta }^k  \circ
                       \sum\limits_{m \ne k} {{\bf A}^m }  = {\bf 0}
                                                           ,\,\,\,\,
             {\bf D}^ +  {\bf A}^k  + {\bf \Theta }^k  = {\bf 0},\quad k =
                                                       1,...,{\rm N}
\end{equation}
\begin{equation}\label{(11.2)}
{\bf D}^ +  {\bf A}^m  \circ {\bf A}^k  + {\bf D}^ +  {\bf A}^k
\circ {\bf A}^m  = 0,\quad k \ne m.
\end{equation}
Суммарное  поле, как  легко видеть (суммируя (\ref{(11.1)}) по
$k$), в  силу (\ref{(11.1)}), является свободным, поскольку,
аналогично механике  взаимодействующих тел, все действующие  силы
внутренние.

Итак взаимодействующие  поля удовлетворяют аналогу второго закона
Ньютона для полей (\ref{(11.1)}), (\ref{(11.2)}),  а для
суммарного заряда-тока выполняется равенство:
\begin{equation}\label{(11.3)}
{\bf D}^ + {\bf \Theta } =  {\bf D}^ + \sum\limits_{m = 1}^M {{\bf
\Theta }^m }  ={\bf 0}.
\end{equation}

Уравнения  суммарного  поля являются следствием  и  дают  первые
интегралы  взаимодействующих полей зарядов-токов. Если в начальный
момент времени заряды-токи  известны,  система  (\ref{(10.1)})
-(\ref{(10.2)}) позволяет определять создаваемые  ими  поля  и  их
совместное изменение во времени  и пространстве.

Рассмотрим  законы  преобразования  энергии полей  при
взаимодействии различных зарядов-токов.

Энергия-импульс для суммарного поля зарядов-токов имеет вид:
\[\displaylines{  {\bf \Xi }_\Theta   = 0,5{\bf \Theta } \circ {\bf \Theta }^*
= 0,5\sum\limits_{k = 1}^N {{\bf \Theta }^k }  \circ \sum\limits_{l = 1}^N
{{\bf \Theta }^{*l} }  = 0,5\left( {\sum\limits_{k = 1}^N {{\bf \Theta }^k
 \circ {\bf \Theta }^{*k} }  + \sum\limits_{k \ne l} {{\bf \Theta }^k
  \circ {\bf \Theta }^{*l} } } \right) =  \cr    = \sum\limits_{k = 1}^N {W_\Theta  ^{(k)}
    + i\sum\limits_{k = 1}^N {P_\Theta  ^{(k)} } }  + {\bf \delta \Xi }_\Theta   \cr}\]
Здесь первое слагаемое - это сумма  энергий-импульсов взаимодействующих зарядов-токов

Введем бикватернион \emph{энергии-импульса взаимодействия}.
 Его действительная часть описывает энергию-импульс взаимодействия одноименных
 зарядов и токов, а мнимая часть - разноименных:
\[
{\bf \delta \Xi }_\Theta   = \delta W_\Theta   + i\delta P_\Theta
= \sum\limits_{k \ne l}^{} {{\bf \Xi }_\Theta  ^{kl} } ,\quad{\bf \Xi
}_\Theta  ^{kl}  = 0,5\left( {{\bf \Theta }^k  \circ {\bf \Theta
}^{*l}  + {\bf \Theta }^l  \circ {\bf \Theta }^{*k} } \right)
\]
$${\bf \Xi }_\Theta  ^{kl}  = {\mathop{\rm Re}\nolimits} \left
( {\rho ^k \rho ^{*l}  + \left( {J^k ,J^{*l} } \right)} \right)
- i\left\{ {{\mathop{\rm Re}\nolimits} \left( {\rho ^k J^{*l}  +
 \rho ^{*l} J^k } \right) + {\mathop{\rm Im}\nolimits} \left[ {J^k ,J^{*l} } \right]} \right\},$$
или в исходных обозначениях:
$${\bf \Xi }_\Theta  ^{kl}  = \frac{{\rho
^{E(k)} \rho ^{E(l)} }}{{\sqrt {\varepsilon _k \varepsilon _l } }}
+ \frac{{\rho ^{(k)H} \rho ^{H(l)} }}{{\sqrt {\mu _k \mu _l } }} +
\sqrt {\mu _k \mu _l } \left( {j^{(k)E} ,j^{(l)E} } \right) +
\sqrt {\varepsilon _k \varepsilon _l } \left( {j^{(k)H} ,j^{(l)H}
} \right) - $$
\[
\displaylines{
   - i\left\{ {\sqrt {\frac{{\mu _l }}{{\varepsilon _k }}}
   \rho ^{(k)E} j^{(l)E}  + \sqrt {\frac{{\varepsilon _l }}
   {{\mu _k }}} \rho ^{(k)H} j^{(l)H}  + \sqrt {\frac{{\mu _k }}
   {{\varepsilon _l }}} \rho ^{(l)E} j^{(k)E}  + \sqrt {\frac{{\varepsilon _k }}
   {{\mu _l }}} \rho ^{(l)H} j^{(k)H} } \right. -  \cr
  \left. { - \sqrt {\varepsilon _k \mu _l } \left[ {j^{(l)E} \,,j^{(k)H} } \right]
  + \sqrt {\varepsilon _l \mu _k } \left[ {j^{(k)E} ,j^{(l)H} } \right]} \right\} \cr}
\]
В результате  получаем условия преобразования энергии при взаимодействии зарядов-токов:
\emph{выделение }энергии, если
$\delta W_\Theta   > 0$;
\emph{поглощение }энергии, если     $\delta
W_\Theta   < 0$;
\emph{сохранение }энергии, если    ${\bf
\delta \Xi }_\Theta = 0$.

\vspace{3mm}

\textbf{ Заключение. }Предложенные здесь  уравнения взаимодействия
ЭГМ-полей,  основаны на гипотезе о магнитном заряде-массе,
симметрирующем уравнения Максвелла. Это позволило назвать такие
поля \textit{электро-грави\-маг\-нит\-ными} и построить законы их
преобразования и взаимодействия, во многом аналогичные законам
Ньютона для материальных тел. Исследование этих уравнений
на инвариантность при преобразованиях Лоренца показало, что
гипотезу об инвариантности закона сохранения заряда-массы для модели ЭГМ-поля,
 рассмотренной в [7,8] принимать нельзя. Здесь показано, что необходимо
введение скалярного поля $a(\tau,x)$ в бикватернион напряженности ЭГМ-поля. Это
 приводит к модификации уравнений Максвелла.

Заметим, что введение $a(\tau,x)$  несколько видоизменит правую часть
уравнения трансформации (\ref{(6.5)}), которую нетрудно выписать,
используя бикватернионную форму этого уравнения (\ref{(6.1)}), вид которой не изменяется.

\vspace{5mm}

 \centerline{  \textbf{Литература}}

\makebox[2em][r]{1. }    Алексеева  Л.А. Гамильтонова форма
уравнений Максвелла  и  ее
     обобщенные решения// Дифференциальные уравнения. Т.39(2003).№6. С.769-776

\makebox[2em][r]{2. }   Алексеева Л.А. Кватернионы гамильтоновой
формы уравнений
    Максвелла// Математический журнал.Т.3.(2003).№4. С.20-24

 \makebox[2em][r]{3. }Ефремов А.П.Кватернионы: алгебра, геометрия и
физические теории//Гиперкомплексные числа в геометрии и
физике.2004,№1(1).С.111-127.

\makebox[2em][r]{4. }  Rastall R.Quaternions in relativity.Rewiew
of modern physics.1964.820-832

\makebox[2em][r]{5. }   Казанова Г. Векторная алгебра.М.1979.120с.

\makebox[2em][r]{6. }   Kassandrov V.V.Buquaternion electroynamics
and Weyl-Cartan
    geometry of space-time//    Gravitation and cosmology.V.1 (1995). №3. P.216-222.

\makebox[2em][r]{7. }   Алексеева Л.А. Об одной модели
электро-гравимагнитного поля.
    Уравнения взаимодействия и законы сохранения// Математический журнал.Т.4.
    (2004).№2. С.23-34.

\makebox[2em][r]{8. }   Алексеева Л.А. Уравнения взаимодействия
А-полей и законы Ньютона  //
 Известия НАН РК. Серия физико-математическая. 2004. №3. С.45-53.

\makebox[2em][r]{9. } Владимиров В.С. Обобщенные функции в
математической физике.М.1976.

\makebox[2em][r]{10. }   Matos C.J., Tajmar M. Advance of Mercury
Perihelion Explained
     by Co\-gra\-vi\-ty/ Advanced Concepts and Studies Officer.2003. E-mail: clovis.de.matos@esa.int

\makebox[2em][r]{11. }   Алексеева Л.А. О замыкании уравнений
Максвелла //  Журнал
    вычислительной математики и математической физики. Т.43(2003).№5.C.759-766.

\end{document}